\documentclass[pra, amsfonts, amssymb, amsmath,doi,reprint, showkeys, twoside,nobibnotes,floatfix, nofootinbib]{revtex4-1}

\usepackage[colorinlistoftodos, color=green!40, prependcaption]{todonotes}
\usepackage{hyperref}

\usepackage{amsthm}
\usepackage{mathtools}
\usepackage{physics}
\usepackage{xcolor}
\usepackage{graphicx}
\usepackage{adjustbox}
\usepackage{placeins}
\usepackage[T1]{fontenc}
\usepackage{lipsum}
\usepackage{csquotes}
\usepackage{textcomp}
\usepackage{url}
\usepackage{graphicx}
\usepackage{amsmath}
\usepackage{bm}
\usepackage{braket}
\usepackage[]{algorithm2e}

\begin{document}

\title{Diagonal and off-diagonal hyperfine structure matrix elements in KCs within the relativistic Fock space coupled cluster theory}

\author{Alexander V. Oleynichenko$^{1,2}$}
\email{aoleynichenko@laser.chem.msu.ru}
\homepage{http://www.qchem.pnpi.spb.ru}
\author{Leonid V. Skripnikov$^{1,3}$}
\author{Andr\'ei Zaitsevskii$^{2,3}$}
\author{Ephraim Eliav$^{4}$}
\author{Vladimir M. Shabaev$^{1}$}

\affiliation{$^{1}$Department of Physics, Saint Petersburg State University, 7/9 Universitetskaya Naberezhnaya, 199034 Saint Petersburg, Russia}
\affiliation{$^{2}$Department of Chemistry, Lomonosov Moscow State University, Leninskie gory 1/3, 119991 Moscow, Russia}
\affiliation{$^{3}$B. P. Konstantinov Petersburg Nuclear Physics Institute of National Research Centre ``Kurchatov Institute'', Gatchina, 188300 Leningrad District, Russia}
\affiliation{$^{4}$School of Chemistry, Tel Aviv University, Tel Aviv, Israel}

\date{\today}

\begin{abstract}

The four-component relativistic Fock space coupled cluster method is used to describe the magnetic hyperfine interaction in low-lying electronic states of the KCs molecule. Both diagonal and off-diagonal matrix elements as functions of the internuclear separation $R$ are calculated within the finite-field scheme. The resulting matrix elements exhibit very weak dependence on $R$ for the separations exceeding 8 \AA , whereas in the vicinity of the ground-state equilibrium the deviation of molecular HFS matrix elements from the atomic values reaches 15\%. The dependence of the computed HFS couplings on the level of core correlation treatment is discussed.

\end{abstract}


\maketitle              % typeset the title of the contribution

\section{Introduction}

Accurate theoretical predictions of properties of the ground and excited states of few-atomic molecules are of crucial importance for preparation and interpretation of different fundamental physical experiments such as laser-synthesis of diatomics~\cite{Ferber2009,Orban2015}, laser cooling of diatomic and polyatomic molecules, experiments aimed at the search for the fundamental symmetry violations effects such as the electron electric dipole moment, etc.
In particular, both diagonal and off-diagonal matrix elements of magnetic dipole hyperfine structure (HFS) and other operators, strongly localized in vicinities of atomic nuclei and sometimes referred to as core property operators, are required for the accurate treatment of non-adiabatic effects~\cite{Petrov:13,Skripnikov:2019a}, parity non-conserving amplitudes~\cite{Geddes:18} and some other transition properties in molecules. In this regard, magnetic dipole hyperfine structure calculations are of particular interest since they are conventionally used as a probe of the accuracy of theoretical predictions.

\begin{figure}
\center
\includegraphics[scale=0.3]{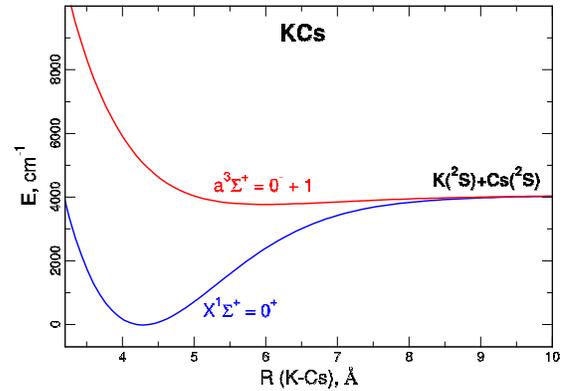}
\caption{Potential energy curves of the $X^1\Sigma^+$ and $a^3\Sigma^+$ states of the KCs molecule.}\label{fig:pec}
\end{figure}

An optimization of paths of laser synthesis of diatomic alkali molecules requires exceptionally complete information on both the energy and radiative properties of their rovibronic states, typically provided by nonadiabatic models including both \textit{ab initio} and spectroscopic data \cite{Pazyuk2015,Klincare2012}.

In this connection the most challenging problem is the accurate description of the complex formed by the ground and the first excited electronic states of alkali dimers, $X{}^1\Sigma^+ \sim a{}^3\Sigma^+$ (see Fig. \ref{fig:pec} for KCs), since rovibrational levels belonging to this complex are exploited as the working ones at the first step of any laser synthesis procedure (photoassociation or magnetoassociation). The $X{}^1\Sigma^+ - a{}^3\Sigma^+$ mixing due to the hyperfine interaction results in extremely complex compositions of rovibronic states near the dissociation limit \cite{Ferber2009,Orban2015}. Most nonadiabatic models of the $X{}^1\Sigma^+ \sim a{}^3\Sigma^+$ complex constructed to date assumed hyperfine interaction matrix elements to be constant with respect to the internuclear distance $R$, and thus their values were derived from the atomic ones (for example, see \cite{Ferber2009} and references therein). However, there are indications that the hyperfine interaction matrix elements actually depend on $R$ at least in the bound region of potential energy curves (PECs) \cite{Docenko2011,Lysebo2013,Temelkov2015}, but the character of this dependence is quite unclear since no highly reliable calculations of the HFS matrix elements in alkali diatomics were reported. The clarification of this question can not only justify the neglect of $R$-dependence for large values of $R$, but also bring the nonadiabatic models to a new level of accuracy which will be adequate to the current level of high-resolution molecular spectroscopy.

In the present paper, we report \textit{ab initio} calculations of the HFS matrix elements in the KCs molecule. The relativistic Fock space coupled cluster (FS-RCC) method \cite{Visscher2001} combined with the denominator-shift technique was employed throughout the paper. This combination has been proven to be a powerful tool for modeling both the energy and radiative properties of such molecules \cite{Zaitsevskii2017,Znotins2019}. Being able to treat relativistic and correlation effects simultaneously, which is of particular importance for really high-precision simulations of systems containing heavy nuclei \cite{Eliav2017}, the FS-RCC method is not well-suitable for analytic calculation of density matrices and calculation of properties due to difficulties with the extraction of explicit wave functions \cite{Szalay1995,Zaitsevskii2018}. This difficulty can be bypassed by using the finite-field (FF) approach. This approach is straightforward for diagonal matrix elements and is based on the Hellmann-Feynman theorem. However, its extension to the case of off-diagonal elements originally introduced in the multireference perturbation theory framework \cite{Zaitsevskii1998} and recently generalized for the case of the Fock space coupled cluster method \cite{Zaitsevskii2018} also allows one to avoid analytic construction of CC transition density matrices. Moreover, within this approach, all off-diagonal matrix elements of interest can be obtained simultaneously. Here we report the first application of the off-diagonal FF technique to calculation of core property matrix elements.
In Sec. \ref{sec:theory} we present the theory underlying the magnetic dipole hyperfine interaction in molecules and the extension of the finite-field technique to off-diagonal HFS matrix elements. Technical details describing the computational approach are given in Sec. \ref{sec:comput}. Sec. \ref{sec:results} reports the calculations of the HFS matrix elements mixing the $X{}^1\Sigma^+$ and $a{}^3\Sigma^+$ states of the KCs molecule, including the discussion of the corresponding selection rules, the role of core relaxation and correlation, the dependence of the HFS matrix elements on internuclear separation and the analysis of possible errors. The final Sec. \ref{sec:conclusion} concludes on our research.

\section{Theoretical considerations}\label{sec:theory}

\subsection{Molecular hyperfine structure}

The magnetic dipole hyperfine interaction of the nuclear magnetic moment $\bm{\mu} = \mu\mu_N \mathbf{I}/I$ (where $\mu$ is the value of the nuclear magnetic moment in nuclear magnetons $\mu_N$ and $\mathbf{I}$ is the nuclear spin) with electrons is given by the following operator (see, e.g.~\cite{Norman2018})
\begin{equation}\label{eq:hhfs}
H_{\rm hf}= \bm{\mu} \cdot\bm{T},
\end{equation}
where the electronic part of the operator is
\begin{equation}\label{eq:hfsEl}
\bm{T}=\sum_i \frac{[\mathbf{r}_i\times \bm{\alpha}_i]}{r_i^3},
%aoley \bm{T}=\sum_i \frac{[\mathbf{r}_i\times \bm{\alpha}_i]}{|\mathbf{r}_i|^3},
\end{equation}
$\bm{\alpha}$ is the set of Dirac alpha matrices and $\mathbf{r}_i$ is a radius-vector of electron $i$ with respect to the center of the chosen nucleus with the magnetic moment $\mu$.

In the present paper we are interested in matrix elements of the operator (\ref{eq:hfsEl}) between two electronic states with wavefunctions $\psi_n$ and $\psi_m$:
\begin{equation}\label{eq:aperp}
A_{\eta}^{nm} = \frac{\mu}{I} \braket{ \psi_n | T_{\eta} | \psi_m},
\end{equation}
where $\eta=x,y,z$.
In these designations the usual ``pa\-ral\-lel'' component of the HFS constant, i.e. diagonal matrix element of the HFS operator, for the given electronic state $\psi_n$ of the diatomic molecule with the total electronic angular momentum projection $\Omega$ onto the internuclear axis $z$ is:
\begin{equation}\label{eq:apar}
A_{||} = \frac{1}{\Omega} A_{\eta}^{nn}=\frac{\mu}{I \Omega} \braket{ \psi_n | T_z | \psi_n}.
\end{equation}

The diagonal matrix elements (\ref{eq:apar}) are non-zero for all $\Omega > 0$; off-diagonal matrix elements are non-zero for electronic states with $\Delta \Omega = 1$ or $0^+$ and $0^-$ states. For the case of the $X^1\Sigma^+ \sim a^3\Sigma^+$ complex in KCs these selection rules result in non-zero matrix elements $1 - 1$, $0^+ - 0^-$, $0^- - 1$ and $0^+ - 1$ (we note that the ${}^3\Sigma^+$ state consists of the $0^-$ and $1$ spin-orbit-split states, while the ${}^1\Sigma^+$ state remains unsplit, $0^+$). At the K(${}^2S$)+Cs(${}^2S$) dissociation limit all these non-zero matrix elements converge to the atomic HFS constant (up to a factor 1/2, arising from the $\Omega = 1/2$ value for the atomic ${}^2S$ state).

It should be emphasized that in all the previous studies of HFS in alkali dimers the non-relativistic approximation to the Hamiltonian (\ref{eq:hhfs}) was employed; the Fermi-contact term was shown to be dominant in the HFS mixing of the $X{}^1\Sigma^+$ and $a{}^3\Sigma^+$ states \cite{Lysebo2013}. The relativistic form of the HFS Hamiltonian (\ref{eq:hhfs}) is more general and precise, including the paramagnetic spin-orbit, Fermi-contact, and spin-dipole hyperfine operators.

\subsection{Finite-field approach to hyperfine interaction matrix element calculations}

Since multireference coupled cluster models do not provide closed expressions for wavefunctions, the evaluation of property matrix elements through constructing pure-state and transition density matrices is normally very costly \cite{Szalay1995} or some linearization approximations are required. However,  most properties represented by one-electron operators can be calculated in a more convenient manner by numerical differentiation with respect to  the amplitude of some external perturbation. This approach usually called the finite-field method is widely used to evaluate expectation values. For example, the magnetic dipole hyperfine structure constant for the electronic state $\psi_n$ can be considered as the partial derivative of molecular energy with respect to the parameter $\lambda$ formally setting the strength of the hyperfine interaction, $H_{\rm hf} \to \lambda H_{\rm hf}$:
\begin{equation}
\braket{\psi_n|T_{\eta}|\psi_n} = \frac{\partial E_n}{\partial \lambda} \left| \begin{array}{l}\\_{\!\lambda=0}\end{array}\right. .
\label{ffdiag}
\end{equation}
The counterpart of Eq.~(\ref{ffdiag}) for off-diagonal matrix elements can be obtained from the approximate Hellmann-Feynman-like relation for effective operators formulated in Ref.~\cite{Zaitsevskii1998}. For the case of the magnetic dipole hyperfine structure, the $\braket{\psi_n|T_{\eta}|\psi_m}$ matrix element between two electronic states with wavefunctions $\psi_n$ and $\psi_m$ satisfy the approximate relation
\begin{equation}
\braket{\psi_n|T_{\eta}|\psi_m} \approx (E_m - E_n) \left< \tilde{\psi}_n^{\perp\perp}(\lambda_\eta) \left| \frac{\partial}{\partial \lambda_\eta} \tilde{\psi}_m(\lambda_\eta)\right. \right> \left| \begin{array}{l}\\_{\!\lambda_\eta=0}\end{array}\right.\!\! .
\label{ffhfs}
\end{equation}
Here $\eta=x,\,y,\,z$ and $\tilde{\psi}^{\perp\perp}$, $\tilde{\psi}$ denote left and right eigenvectors of the field-dependent non-Hermitian FS-RCC effective Hamiltonian acting in the field-independent (constructed for $\lambda=0$) model space, and $E_{m}$ and $E_{n}$ are the field-free energies of the involved states. The derivative in the r.h.s. of Eq. (\ref{ffhfs}) is readily estimated  by the finite difference method. Although the formula (\ref{ffhfs}) involves only the effective Hamiltonian eigenvectors (the model space projections of many-electron wavefunctions), the resulting matrix elements estimates implicitly incorporate the bulk of the contributions from the remainder parts of these wavefunctions~\cite{Zaitsevskii1998,Zaitsevskii2018}. 

Note that the approximate matrix elements $\braket{\psi_n|T_{\eta}|\psi_m}$ and $\braket{\psi_m|T_{\eta}|\psi_n}^*$ defined by Eq.~(\ref{ffhfs}) generally do not coincide due to the non-Hermitian nature of the FS-RCC effective Hamiltonian. To prevent the appearance of unphysical differences between these quantities (which are normally within 1\%), the preliminary transformation of the non-Hermitian FSRCC effective Hamiltonians to the Hermitian form \emph{via} the symmetric orthogonalization of their eigenvectors can be performed.

For the case of the KCs molecule the differentiation (\ref{ffdiag}) allows one to obtain only the diagonal $1 - 1$ matrix element; the other three ($0^+ - 0^-$, $0^- - 1$ and $0^+ - 1$) can be calculated within the relation (\ref{ffhfs}). Furthermore, in alkali dimer molecules the lowest $0^-$ and $1$ electronic states corresponding to the $a^3\Sigma^+$ state are nearly not perturbed by the spin-orbit interaction with higher-lying states and hence are actually degenerate (see Fig. \ref{fig:pec}). This results in insurmountable numerical instabilities when using (\ref{ffhfs}) and the $0^- - 1$ matrix element cannot be accessed within the current formulation of the finite-field technique. However, this does not give rise to any problem since the $1 - 1$ and $0^- - 1$ matrix elements are expected to have nearly the same values (and the same dependence on internuclear distance, which is even more important here).

\section{Computational details}\label{sec:comput}

The relativistic Fock space coupled cluster method with single and double excitations (FS-RCCSD), recently demonstrated to be ideally suitable for high-precision modeling of energy and radiative properties of alkali diatomics \cite{Znotins2019}, was used throughout.
The coupled cluster calculations were carried out within the EXP-T program package \cite{Oleynichenko2020}
\footnote{See also http://www.qchem.pnpi.spb.ru/expt, accessed on 28 June 2020}.
Molecular spinors and molecular integrals, including the HFS ones, were calculated within the DIRAC code \cite{DIRAC17,Saue2020}.

Since the magnetic dipole hyperfine interaction is mainly determined by the behavior of electronic wavefunctions in the vicinity of the atomic nucleus, the all-electron four-component relativistic calculations with the Dirac-Coulomb Hamiltonian were performed in order to achieve high and controllable accuracy. An alternative two-step approach \cite{Titov2005} based on a restoration of spinors in the vicinity of a heavy nucleus is formally less precise (however, it can be used in combined schemes to treat high-order correlation effects, basis set corrections, etc. to increase the final accuracy ~\cite{Skripnikov:16b}). 
%ls: ошика восстановления примерно 7%. Всё же, я бы не упоминал только минусы этой схемы, т.к. есть и плюсы. Например, можно использовать сматые базисы и смотреть CCSDT(Q). В полноэлектронные технически пока сжатые базисы не использовать; т.е. свобода действий в полноэлектронных технически куда ниже. Также в двухшаговой схеме можно отключить СО ТОЛЬКО ДЛЯ ВАЛЕТНЫХ (которая для некоторых свойств не очень-то и важна) и рассматривать базисы, например, с 15G, 15H, 15I функциями и более на уровне CCSD(T). В полноэлектронных для таких расчётов не хватит и ЦОДа.
%(with typical errors of order 15-20\%)). 
%end ls
In the present calculations, all electrons were correlated; the role of core relaxation and correlations is discussed below in Sec. \ref{sec:results}.

The ground state of KCs${}^{2+}$ was used as the reference state in the FS-RCC calculations and the electronic states of the neutral KCs molecule were constructed in the $(0h,2p)$ Fock space sector (two particles over the Fermi vacuum). The active space comprised 12 lowest Kramers pairs of spinors with the total angular momentum projection $|m_j| = 1/2$ , 6 pairs with $|m_j| = 3/2$ and 2 pairs with $|m_j| = 5/2$. At the dissociation limit this corresponds to the $ns$, $np$, $(n-1)d$ and $(n+1)$s spinors of the isolated K and Cs atoms. In order to suppress numerical instabilities due to intruder states in the FS-RCC calculations, the new modification of denominator shift technique \cite{Zaitsevskii2017} based on the simulation of an imaginary shift was used (see Appendix A).  
We have used flexible uncontracted basis sets $(18s14p6d1f)$ for K and $(28s22p15d5f)$ for Cs (see Supplementary \cite{Supplementary}). These sets are based on the Dyall's basis set family \cite{Dyall2009}\footnote{Available from the Dirac web site, http://dirac.chem.sdu.dk, accessed on 28 June 2020.} and augmented with functions required to reproduce the atomic HFS constants and low-lying energy levels with the fairly good accuracy (see Table \ref{tab:athfs}).
The finite-field approach with the 
%az parameter $\lambda = 10^{-8}$ MHz${}^{-1}$ ($\approx 1.3 \cdot 10^{-4}$ a.u.)
numerical differentiation step $\Delta\lambda = 10^{-8}$ MHz${}^{-1}$ ($\approx 1.3 \cdot 10^{-4}$ a.u.) was employed to calculate hyperfine interaction expectation and transition matrix elements; this value was found to be optimal for ensuring numerical stability.

\begin{table}[h]
\center
\def\arraystretch{1.2}
\caption{Atomic hyperfine structure constants and energy levels of $^{39}$K and $^{133}$Cs calculated within the FS-RCCSD method and compared with experimental data.}\label{tab:athfs}
\begin{tabular}{ccrrcrr}
\hline
\multicolumn{2}{c}{State} & \multicolumn{2}{c}{A, MHz} & & \multicolumn{2}{c}{
Transition energy, cm${}^{-1}$} \\
\cline{3-4} \cline{6-7}
& & FS-RCCSD & exptl\cite{Tiecke2011,Steck2010} & & FS-RCCSD & exptl\cite{Sansonetti2005} \\
\hline

\multicolumn{7}{c}{\it Potassium} \\

$4s$ & ${}^2S_{1/2}$ & 222.4   & 230.9 & & 0     & 0     \\
$4p$ & ${}^2P_{1/2}$ & 26.7     & 27.8 & & 12966 & 12985 \\
     & ${}^2P_{3/2}$ & 6.0      &  6.1 & & 13025 & 13043 \\

\multicolumn{7}{c}{\it Cesium} \\

$6s$ & ${}^2S_{1/2}$ & 2252.7 & 2298.2 & & 0     & 0     \\
$6p$ & ${}^2P_{1/2}$ & 280.3   & 291.9 & & 11184 & 11178 \\
     & ${}^2P_{3/2}$ & 48.6     & 50.3 & & 11732 & 11732 \\
\hline
\end{tabular}
\end{table}

We considered only the most abundant isotopomer ${}^{39}$K${}^{133}$Cs; the HFS constants for other isotopes can be obtained simply by scaling. The qualitative picture of energy levels at the K(${}^2S$)+Cs(${}^2S$) dissociation limit is strongly dominated by the HFS induced by the ${}^{133}$Cs nucleus ($I$ = 7/2, $A_{6s ^2S_{1/2}}$ = 2298.1579425 MHz \cite{Steck2010}); the effect induced by the ${}^{39}$K nucleus ($I$ = 3/2, $A_{4s ^2S_{1/2}}$ = 230.8598601(3) MHz \cite{Tiecke2011}) is much smaller. The picture is actually the same for the two other natural potassium isotopes (${}^{40}$K and ${}^{41}$K). Since the magnetic dipole hyperfine interaction is a one-electron property essentially defined by the behavior of the wavefunctions in the vicinity of nuclei, the HFS matrix element dependencies $A(R)$ can be calculated for the K and Cs nuclei separately. The contributions to the hyperfine structure induced by nuclear electric quadrupole moments are negligible for alkali diatomics \cite{Aldegunde2008,Lysebo2013} and hence are not considered here. The values of nuclear dipole magnetic moments, $\mu_{^{39}K}$ = 0.39147$\mu_N$ and $\mu_{^{133}Cs}$ = 2.582025$\mu_N$, were taken from Ref.~\cite{Stone2005}.

Special attention is to be paid to the accuracy limitations of the model employed. The most important factors restricting the overall accuracy of the calculated HFS matrix elements are the following ones:
(i) the approximation to the re\-la\-ti\-vis\-tic Hamiltonian (lack of Breit and QED terms),
(ii) restrictions of the point dipole model of nuclear magnetization distribution,
(iii) the approximation to the cluster operator (e.g. CCSD) used in the FS-RCC calculation and
(iv) errors caused by the basis set incompleteness.
The latter factor can be suppressed rather efficiently by using extended uncontracted basis sets designed specifically for HFS calculations. The contribution of the Breit two-electron interaction to the $A_{6s\ ^2S_{1/2}}$ constant of Cs atom was found to be less than 0.5\% and even smaller for lighter alkali atoms \cite{Safronova1999}, thus the Dirac-Coulomb approximation can be regarded as precise enough. Errors arising from the point-dipole approximation were estimated recently to be smaller than 1\% \cite{Ginges:2018,Prosnyak2020}. The effect of higher-order excitations in the cluster operator was found to be small for the Cs $6s\ ^2S_{1/2}$ HFS constant (ca.~$\sim$2\%) \cite{Haase2020}. Finally, errors due to non-compensated denominator shifts used to suppress intruder states in a wide range of $R$ are negligible for such low-lying electronic states \cite{Zaitsevskii2017,Znotins2019}. However, the clarification of the $R$-dependence of HFS matrix elements does not require extremely high accuracy; an error of about 5\% is to be regarded as quite satisfactory.

\section{Results and discussion}\label{sec:results}

\begin{figure}[h]
\center
\includegraphics[width=0.49\textwidth]{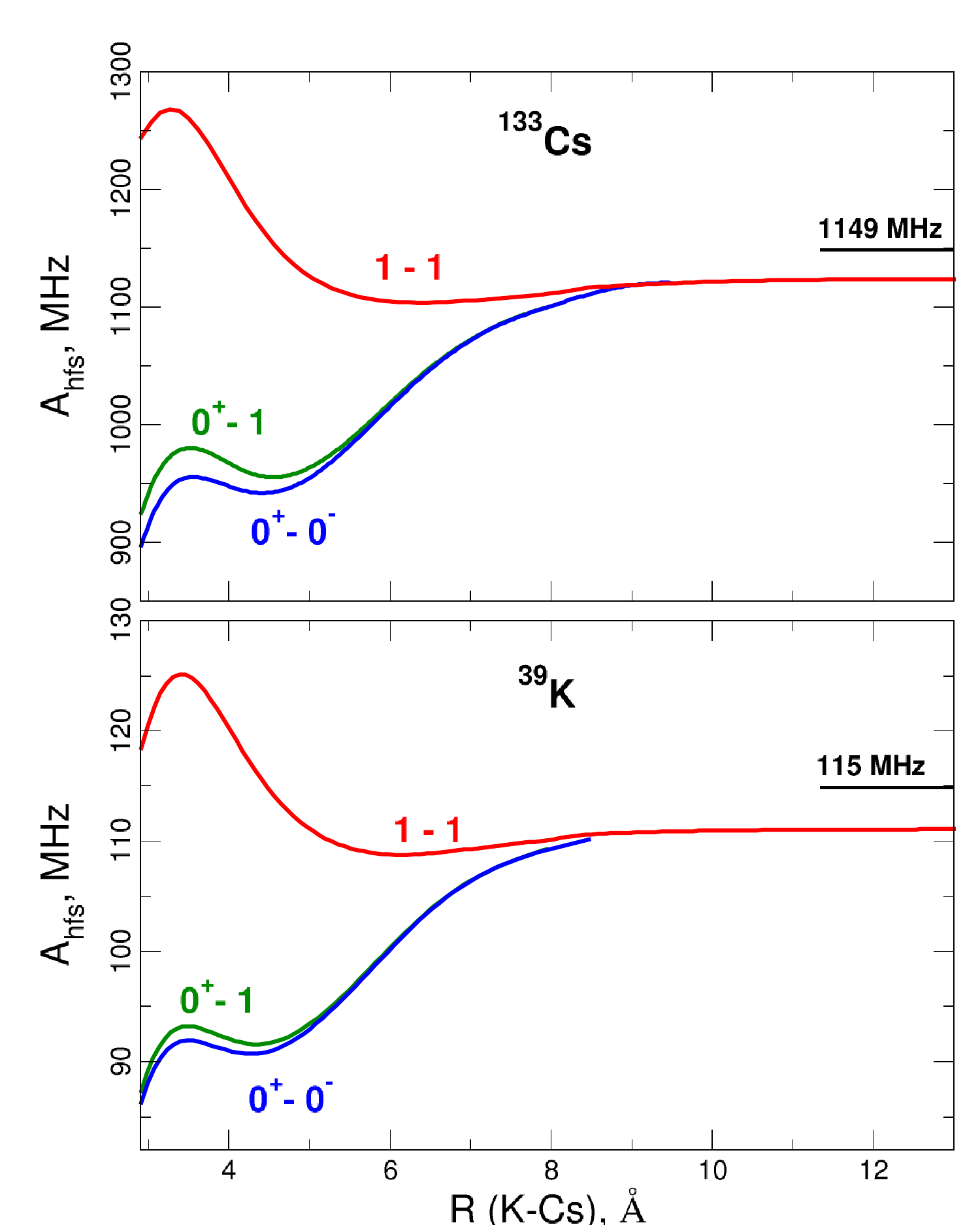}
\caption{Hyperfine coupling matrix elements mixing the $X^1\Sigma^+_{0^+}$ and $a^3\Sigma^+_{0^-,1}$ states of ${}^{39}$K${}^{133}$Cs for the HFS induced by the ${}^{133}$Cs nucleus (top panel) and the ${}^{39}$K nucleus (bottom panel). The black horizontal ticks show the corresponding experimental values for the atomic hyperfine structure constant $A_{at}/2$.}\label{fig:hfs}
\end{figure}

The magnetic dipole hyperfine interaction matrix elements between the $X^1\Sigma^+_{0^+}$ and $a^3\Sigma^+_{0^-,1}$ states of ${}^{39}$K${}^{133}$Cs as functions of the internuclear distance are shown in Fig. \ref{fig:hfs}. The obtained $R$-dependencies of the $0^-$ -- $0^+$ coupling elements closely resemble those reported in \cite{Lysebo2013} for the Fermi-contact interaction in the $a^3\Sigma^+_u$ state of homonuclear alkali dimers.

The widely used assumption that the hyperfine coupling matrix elements are constant is not correct for the ``bonding'' region of potential curves ($R <8$ \AA) where the deviations from the corresponding atomic values reach 15\% (see Fig. \ref{fig:pec}). However, these deviations are small for a larger $R$. Thus the variation of the matrix elements of the magnetic dipole hyperfine interaction with $R$ is not expected to affect significantly the hyperfine mixing near the dissociation limit.
This conclusion is in agreement with the experimentally observed hyperfine splittings which are nearly constant with respect to the vibrational quantum number for the $a^3\Sigma^+$ state (see Fig. 5 in \cite{Ferber2009}).
This fact legitimizes the use of atomic values for the modeling of laser synthesis \textit{via} Feshbach resonances or photoassociation.

The variations of HFS matrix elements functions are essentially confined to the ``bonding'' domain of $R$ values. It is thus natural to suppose that these variations originate mainly from the redistribution of valence-shell electrons during the formation of chemical bonds. 
In other words, these variations can be considered as ``chemical shifts''. This feature of the HFS matrix elements can be used to reduce the number of adjustable parameters in non-adiabatic models considering heteronuclear alkali dimers.

\begin{figure}[h]
\center
\includegraphics[width=0.49\textwidth]{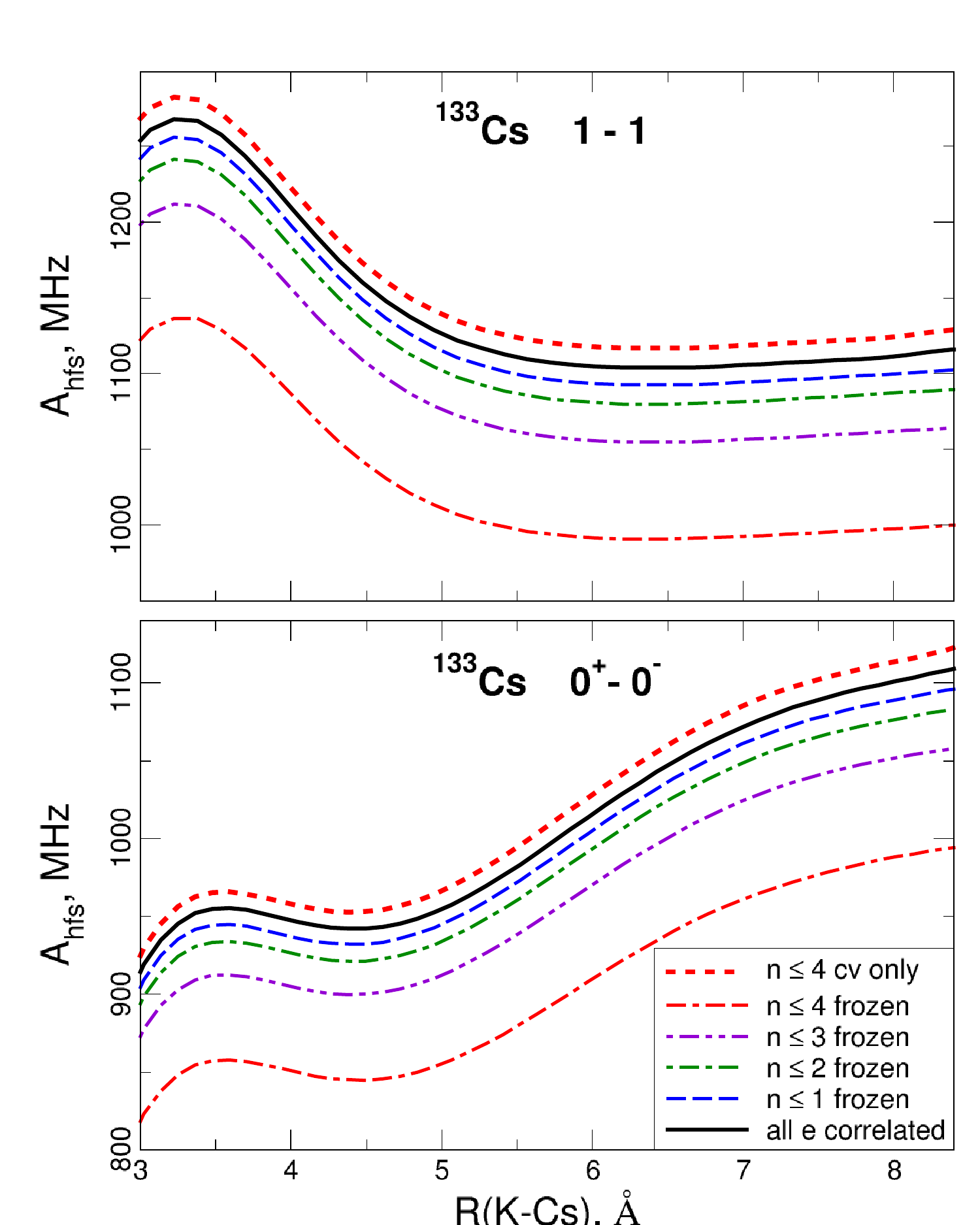}
\caption{Hyperfine coupling matrix elements for the HFS induced by the ${}^{133}$Cs nucleus obtained at the different levels of treatment of core electrons. Black solid line: all electrons are correlated; ``frozen'': core shells with the principal quantum numbers $n$ are completely excluded from correlation treatment; dotted line, ``$n \le 4$ cv only'': only core-valence correlations and relaxation effects for the $n \le 4$ core shells of Cs are incorporated (the corresponding results for the smaller Cs cores are practically indiscernible from those of the full correlation treatment). }
\label{fig:core}
\end{figure}

The elucidation of the role of core relaxation and correlations is of great interest since the neglect of these effects can in principle lead to significant computational savings, making feasible calculations of core properties for large systems. Within the FS-RCCSD approach, core correlations are associated with the double excitations in the cluster operator which involve two core spinors. Omitting these excitations, one neglects the core correlations but retains the bulk of the core relaxation and core-valence correlation which are described by cluster terms involving only one core spinor. Fig.~\ref{fig:core} presents the $R$-dependencies of the $1 - 1$ and $0^+ - 0^-$ matrix elements of the Cs-induced HFS calculated with different sizes of the Cs core. It can be seen that electron correlations restricted to the core including all spinors with the principal quantum numbers $n \le 3$ practically do not contribute to the HFS matrix elements. On the contrary, the results are distorted significantly if the core relaxation and core-valence correlations are omitted as well (i.e. all core spinors are frozen at the post-SCF correlation calculation step): the deviation from the results with the full correlation treatment (black solid curve) reaches 5\% for the frozen $n \le 3$ core and 10\% for the frozen $n \le 4$ core. The latter error is of the same order of magnitude as that of the two-step non-variational core restoration scheme \cite{Titov2005}, which also completely neglects core relaxation and core-valence correlation contributions. However, since these contributions are actually independent from the chemical environment (see Fig. \ref{fig:core}), it might be reasonable to keep the core spinors frozen in molecular calculations and to extract the missing contributions from atomic calculations. A similar scheme can be used for evaluating other core properties. Furthermore, HFS matrix elements calculated as functions of molecular geometry with the frozen core can be adjusted to experimental atomic values simply by appropriate shift.

%az Actually the same general patterns can be expected for hyperfine interactions mixing electronic states converging to dissociation limits other than ${}^2S+{}^2S$. Note that since the atomic HFS constants are an order of magnitude smaller for the $P$-states, the account for variation of hyperfine couplings for the corresponding molecular electronic states within non-adiabatic models unlikely to be really required at the current state of the molecular spectroscopy capabilities, especially for the lighter alkali dimers.

\section{Conclusions}\label{sec:conclusion}

The relativistic four-component Fock space coupled cluster method was used to calculate both diagonal and off-diagonal matrix elements of the magnetic dipole hyperfine interaction operator within the finite-field approach. The numerical stability of the solutions of the FS-RCC equations was ensured by the imaginary shift simulation technique. From the results for the dissociation limit, the overall error of our estimates should be about 3-5\%.

Our results demonstrate that the HFS coupling matrix elements in KCs are nearly constant for large and intermediate internuclear distances ($R > 8$ \AA, i.\ e. including the area of small separations of the $X^1\Sigma^+$ and $a^3\Sigma^+$ states in energy). This finding justifies the use of unscaled atomic hyperfine structure data for modeling the Feshbach resonances in alkali diatomics. For smaller $R$ the variation of HFS matrix elements become significant; the deviation of these entities from the corresponding atomic values reaches ca. 15\% in the vicinity of equilibrium distances for the ground and strongly bound excited states. A detailed study of the effect of the core relaxation, core-core and core-valence electron correlations on the computed matrix elements was performed. Core-core correlations practically do not contribute to the matrix elements under study. Rather significant (up to $\sim 10$\%) core relaxation and core-valence correlation contributions were found nearly independent on $R$. Thus the essential information on the variation of the HFS matrix elements upon chemical bond formation (``chemical shift'') can be obtained via relatively inexpensive calculations within the frozen core approximation. 

%az To authors' knowledge, the calculations performed are the most comprehensive and high-level \textit{ab initio} calculations of the HFS dependence on molecular geometry, at least for the alkali dimer molecules. The \textit{ab initio} HFS matrix elements can be calculated in the frozen core approximation and then properly adjusted to experimental HFS constants for atoms. HFS coupling functions obtained in such manner are \textit{a priori} better in quality than simply the $R$-independent constants, the use of such functions can result in the great improvement of accuracy of non-adiabatic models (OF WHAT?).
%ls в нашей работе 10.1103/PhysRevA.88.010501 недиагональные матричные элементы вычислялись для учёта неадиабатических поправок к HFS

%az (((I would like to omit:))) Further developments are required to construct an effective-potential-like model taking into account the core relaxation and core-valence correlation effects, which were demonstrated to be inevitable for cheap, but really high-precision calculations of core properties.

\section{Acknowledgements}

Authors are grateful to E. A. Bormotova and A. V. Stolyarov for fruitful discussions.

This work has been carried out using computing resources of the federal collective usage centre Complex for Simulation and Data Processing for Mega-science Facilities at NRC ``Kurchatov Institute'', \url{http://ckp.nrcki.ru/}, and computers of Quantum Chemistry Lab at NRC ``Kurchatov Institute'' -- PNPI.

The reported study was funded by RFBR according to the research project \textnumero19-33-50122\textbackslash19.

\section{Appendix A}

Although the strategy of solving the relativistic many-electron problem generally followed that described in Refs.~\cite{Zaitsevskii2017,Zaitsevskii2018a}, we employed a somewhat different scheme to suppress numerical instabilities which is described below.

Due to the model space completeness, the straightforward application of the conventional FS-RCC formulation in wide ranges of internuclear separations is blocked by numerical instabilities caused by intruder states (see e.g. Refs.~\cite{Evangelisti1987,Zaitsevskii2017} for detailed discussions). The presence of intruder states manifests itself as the appearance of poles of the $D_K^{-1}$ factors arising from small or zero energy denominators in FS-RCC amplitude equations at certain nuclear geometries ($D_K$ stands for the energy denominator for the excitation $K$ \cite{Zaitsevskii2017}).

Similarly to the technique used in our previous work~\cite{Zaitsevskii2017}, we replace the original energy denominators $D_K$ in the FS-RCC amplitude equations by their shifted counterparts in such a way that numerical instabilities are suppressed without significant distortions of the description of low-lying states. However, the particular Ans\"{a}tze for the dependencies of real shifts on the original denominator values proposed in Refs.~\cite{Eliav2005,Zaitsevskii2017,Zaitsevskii2018a} can give rise to non-negligible systematic errors and rather strong dependencies of results on the particular choice of shift parameters. Though this problem is efficiently solved through extrapolating to the zero shift limit \cite{Eliav2005,Zaitsevskii2018a}, a less cumbersome approach is desirable for molecular excited state calculations in wide areas of nuclear configurations.  

To construct a potentially better Ansatz, let us turn to the idea of intruder state avoidance by means of imaginary denominator shifts~\cite{Surjan1996,Forsberg1997}. The imaginary analog of the denominator shift proposed in Ref.~\cite{Zaitsevskii2017} would imply the replacement
\begin{equation}
D_K\;\Longrightarrow\;D^C_K(m)=D_K+s_K\,
\left(\displaystyle\frac{|s_K|}{\left| D_K+s_K i \right|}\right)^{m} i,
\label{imags}
\end{equation}
where the $s_K$ parameters (shift amplitudes) are real; these parameters are supposed to have the same value for all classes of excitations. The non-negative integer $m$ (attenuation parameter) is common for all excitations $K$. The imaginary part of $D_K^C$ tends to $s_K$ for nearly zero denominators $D_K$ and becomes small for ''well-defined'' (large negative or positive) $D_K$. The larger the $m$ value, the smaller the shifting of ``well-defined'' energy denominators. The essential advantage of imaginary shifts compared to their real counterpart introduced in Ref.~\cite{Zaitsevskii2017} arises from the fact that $D_K^C(m)^{-1}$ with any non-zero $s_K$ have no poles on the real axis regardless the behaviour of $D_K$ as functions of the nuclear geometry. Note, however, that the straightforward application of (\ref{imags}) could lead to non-physical imaginary contributions to effective Hamiltonian eigenvalues; furthermore, one would lose the possibility to perform FS-RCC calculations of diatomic (or other highly symmetric) molecules in real numbers. Therefore instead of using Eg.~(\ref{imags}), we replace the original energy denominators $D_K$ in the FS-RCC amplitude equations by real shifted denominators $D^{\prime}_K(m)$
\begin{equation}
D_K\;\Longrightarrow\;D^{\prime}_K(m)=D_K + \frac{s_K^2}{D_K}\left(\displaystyle\frac{s_K^2}{  D_K^2+s_K^2}\right)^{m}
\label{imagreal}
\end{equation}
satisfying the requirement 
\begin{equation}
%D_K\;\Longrightarrow\;D_K^C:\quad 
\frac{1}{D^{\prime}_K(m)}={\rm Re}\left(\frac{1}{D_K^C(m)}\right)
\end{equation}
and thus simulating in a sense the modification (\ref{imags}) without leaving the real axis. 

The modification (\ref{imagreal}) is a straightforward generalization of the ``imaginary shifting'' widely used in second-order multireference many-body perturbation theories \cite{Forsberg1997,Witek2002}. Assuming $m=0$ in Eq.~(\ref{imagreal}), one immediately arrives at the expression for energy denominators introduced in the cited works. The increase of $m$ should reduce the errors introduced through the replacement of original denominators by shifted ones but can slow down (and finally ruin) the convergence of the iterative solution of the modified cluster equations in the presence of intruder states. A series of single-point calculations demonstrated that the use of a larger $m$ ($m\ge{}2$) is essential for suppressing the dependence of the resulting energies on the choice of shift amplitudes $s_K$. All results presented in Sec.~\ref{sec:results} were obtained with shift parameters $s=-0.2$ a.u. and $m=3$ in the $(0h,2p)$ Fock space sector; FS-RCC equations for the $(0h,0p)$ and $(0h,1p)$ were not modified.

\bibliography{hfs}

\end{document}